\documentclass[pra,twocolumn,groupedaddress,showpacs,byrevtex]{revtex4}
\usepackage{graphics,amssymb,amsmath}

\newcommand{\sect}[1]{Sec.~\ref{#1}}
\newcommand{\fig}[1]{Fig.~\ref{#1}}
\newcommand{\eqn}[1]{Eq.~(\ref{#1})}
\newcommand{\eqns}[2]{Eqs.~(\ref{#1}) and (\ref{#2})}
\newcommand{\ket}[1]{| #1 \rangle}

\newcommand{\braket}[2]{\langle #1 | #2 \rangle}

\newcommand{\be}{\begin{eqnarray}}
\newcommand{\ee}{\end{eqnarray}}

\newcommand{\intd}[1]{\int \!\! d #1 \ }
\newcommand{\wt}[1]{\widetilde{#1}}

\begin{document}

\title{Preparing encoded states in an oscillator}

\author{B.~C.~Travaglione}
 \email{btrav@physics.uq.edu.au}
\author{G.~J.~Milburn}
 \affiliation{
  Centre for Quantum Computer Technology, University of Queensland,
  St. Lucia, Queensland, Australia
  }

\date{May 18, 2002}

\begin{abstract}

Recently a scheme has been proposed for constructing quantum error-correcting
codes that embed a finite-dimensional code space in the infinite-dimensional
Hilbert space of a system described by continuous quantum variables.
One of the difficult steps in this scheme is the preparation of the encoded
states.
We show how these states can be generated by coupling a continuous quantum
variable to a \emph{single} qubit.
An ion trap quantum computer provides a natural setting for a continuous
system coupled to a qubit.
We discuss how encoded states may be generated in an ion trap.

\end{abstract}

\maketitle

\section{Introduction}

It appears, in principle, that the laws of quantum mechanics allow certain
mathematical problems to be solved more rapidly than can be done using a
classical computer \cite{Nielsen00,Preskill98}.
However, in order to accomplish this task, the state of a quantum system must
maintain coherence, despite unwanted interactions with the environment.
There have been a number of proposed mechanisms for protecting quantum
information during a computation 
\cite{Shor95,Steane96,Knill96,Chau97,Rains97,Gottesman96,Calderbank97}.
Recently, it has been shown \cite{Gottesman01} that a $d$-dimensional quantum 
system
(here we only consider $d=2$) can be embedded in an infinite-dimensional
Hilbert space, such that a universal set of fault-tolerant quantum gates can
be implemented using linear optical operations, squeezing, homodyne detection,
and photon counting. The qubits are embedded in the continuous system in a
manner which protects the quantum information against small shifts in the
canonical quantum variables, $q$ and $p$.
Ideally, the encoded states are an infinite sum of delta
functions in both $q$ and $p$. 
Of course, such states are non-normalizable, and unphysical. Hence they must be
approximated.
It has been proposed \cite{Gottesman01} that these approximate encoded states 
could be generated by a procedure involving a 
non-linear interaction Hamiltonian of the form,
\be\label{mirror}
 H^\prime &\propto& qb^\dagger b,
\ee
where $q$ is the position operator of one variable, and $b$
($b^\dagger$) is the annihilation (creation) operator of a second variable.
Unfortunately, interactions of the form given in 
\eqn{mirror} have proven very difficult to implement. They generally require
the radiation pressure of photons to move a 
macroscopic object (a mirror) \cite{Giovannetti00}.

Here we show that approximate encoded states can be generated by
coupling the continuous variable to a \emph{single} qubit, and performing a
sequence of
operations similar to a quantum random walk algorithm \cite{Travaglione02}.

In \sect{ideal}, we briefly review the continuous variable encoding scheme 
proposed in Gottesman et al. \cite{Gottesman01}. 
In \sect{approx} we show how approximate encoded states can be
non-deterministically generated by coupling the continuous variable to a qubit. 
We then discuss in \sect{errorstates} how error recovery can be performing by
deterministically preparing ancilla variables.
Finally, in \sect{iontrap} we discuss how
an ion trap quantum computer could be used to generate approximate
encoded states, and therefore provide an important proof of principle.

\section{Encoding a qubit in an oscillator}\label{ideal}

Quantum computation is generally formulated in terms of interacting two level
quantum systems, or qubits. The choice of two level quantum systems is
partially because it is easy to draw analogies with the classical bit, but
also because a two level system is the simplest non-trivial system; and
increasing the number of levels only increases the computation efficiency by 
a constant of proportionality.

However, with the goal of building a quantum computer in mind, two level
quantum systems are by no means the most natural choice. Most physical systems,
even in their most elemental form, are represented by many more than two
levels. Indeed, many quantum systems are naturally described by a continuous
variable (infinite dimensional Hilbert space).
Such continuous quantum systems have been well
studied, and proposals have been made for performing analog quantum
computation using such systems \cite{Braunstein98b,Lloyd98,Lloyd99}. 

\subsection{Ideal Encoded States}

\begin{figure}[ht]
 \centering
 \scalebox{0.75}{\includegraphics{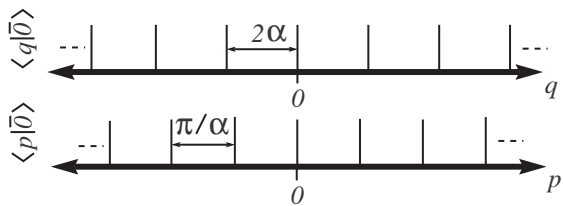}} \\ (a) 
 \scalebox{0.75}{\includegraphics{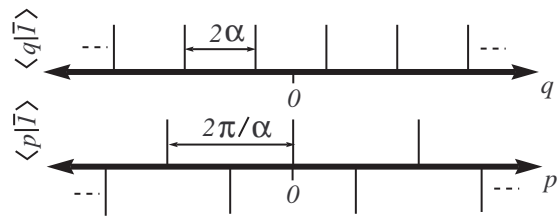}} \\ (b)
 \caption{(a) Ideal wave function, in both position and momentum, of the
 encoded zero state, $\ket{\bar{0}}$. In position space, the wave-function is 
 an infinite sum of delta functions, separated by $2 \alpha$, in momentum space,
 the wave-function is an infinite sum of delta functions separated by 
 $\pi/\alpha$. (b) Ideal wave functions of the encoded one state,
 $\ket{\bar{1}}$.}
 \label{ideal01}
\end{figure}

Gottesman et al. \cite{Gottesman01} discuss how to embed a qubit in a continuous
quantum system, so that the extra degrees of freedom within the system can be 
used to correct errors which arise from unwanted interactions with the 
environment.
Ideally, an encoded zero state, $\ket{\bar{0}}$, will be represented in position
space by the wave function
\be\label{zeroq}
 \braket{q}{\bar{0}} &=& 
   \sum_{s=-\infty}^{\infty} \!\! \delta(q-2\alpha s)
 =  \sum_{s=-\infty}^{\infty} \! e^{i\pi sq/\alpha},
\ee
and thus in momentum space, it has the wave function 
\be\label{zerop}
 \braket{p}{\bar{0}} &=& 
   \sum_{s=-\infty}^{\infty} \!\! \delta(p-\frac{\pi s}{\alpha})
 =  \sum_{s=-\infty}^{\infty} \! e^{i2sp\alpha}.
\ee
Whilst the encoded one state, $\ket{\bar{1}}$, is represented in position
and momentum space by the wave functions,
\be
 \braket{q}{\bar{1}} &=&
  \sum_{s=-\infty}^{\infty} \!\! \delta(q-2\alpha(s-\frac{1}{2}))
 =  \sum_{s=-\infty}^{\infty} \! e^{i\pi(\frac{sq}{\alpha}-1)} \label{oneq}\\
 \braket{p}{\bar{1}} &=&
 \sum_{s=-\infty}^{\infty} \!\! (-1)^s \delta(p-\frac{\pi s}{\alpha})
 =  \sum_{s=-\infty}^{\infty} \! e^{i(2s-1)p\alpha}. \label{onep}
\ee 
The wave functions for the encoded zero state are depicted in \fig{ideal01}(a),
whilst \fig{ideal01}(b) depicts the wave functions for the encoded one
state.
Clearly the zero and one encoded states are orthogonal,
\be
 \braket{\bar{0}}{\bar{1}} &=& 0.
\ee

\subsection{Error Recovery}

For the details of how quantum computation is performed with these encoded
states we direct the reader to Gottesman et al. \cite{Gottesman01}. Here we
review the error recovery procedure, which protects these encoded states
against shifts in position, $q$, and momentum, $p$, of size 
\be\label{shifts}
  |\Delta q| < \frac{\alpha}{2} \quad &\mathrm{and}& \quad
  |\Delta p| < \frac{\pi}{2 \alpha}.
\ee
Suppose we have an encoded qubit in some arbitrary superposition of zero and
one,
\be\label{super}
 \ket{\psi}_e &=& c_0 \ket{\bar{0}} + c_1 \ket{\bar{1}}.
\ee
Assume an error occurs to the state $\ket{\psi}_e$, such that the wave function 
is shifted in the position variable by some amount $\epsilon < \alpha/2$. We 
wish to correct this error without destroying the state. This can be 
accomplished by using an ancilla variable, prepared in the state 
\be\label{ancilla}
 \ket{\phi}_a &=& (\ket{\bar{0}} + \ket{\bar{1}})/\sqrt{2},
\ee 
and an interaction Hamiltonian of the form
\be\label{Herrorq}
 H_1 &=& q_e p_a,
\ee
where the subscript $e$ denotes the encoded qubit variable, and the subscript 
$a$ denotes the ancilla variable. After the two systems have interacted, we can 
measure the $q$ variable of the ancilla system, which will allow us to determine
the value of $\epsilon$. This error can then be corrected by applying an 
appropriate displacement operation to the encoded qubit system.
Likewise, a shift of $\epsilon < \pi/2\alpha$ in the momentum variable can be 
corrected using an ancilla system prepared in the $\ket{\bar{0}}$ state, and
evolving according to the interaction Hamiltonian, 
\be\label{Herrorp}
 H_2 &=& p_e q_a.
\ee

\section{Preparing encoded states using a qubit}\label{approx}

Once prepared, it is hoped that the error recovery
procedure will be able to maintain the encoded states.
However, preparation of the encoded states is not trivial. As has already been
stated, we can only prepare approximate encoded states. In this section we 
show how approximate encoded states can be prepared with the aid of a single
ancilla qubit. Our preparation scheme is non-deterministic, in that a valid
approximate encoded state will only be prepared with some probability less than
one, however, we will know when our preparation procedure has worked.

\begin{figure}[ht]
 \centering
 \scalebox{0.70}{\includegraphics{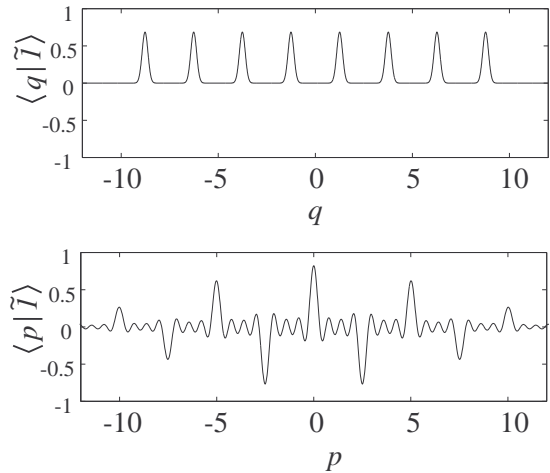}}
 \caption{Wave-function, in both position and momentum, of the 
 approximate
 encoded zero state, $\ket{\widetilde{1}_3}$. This approximate encoded state
 will be generated with probability 1/8, by first squeezing the continuous
 variable in
 momentum quadrature, and then applying the sequence of operations and
 measurements described in the text.}
 \label{approx1}
\end{figure}

We shall denote approximate encoded zero and one states with the symbols 
$\ket{\wt{0}}$ and $\ket{\wt{1}}$. As in \cite{Gottesman01}, we begin 
the preparation procedure with the quantum system in the ground  state of the
oscillator, $\ket{0}$, and apply squeezing in the $q$ quadrature. 
This creates the state
\be\label{Sstate}
 \braket{q}{s} &=& g(q,\Delta),
\ee
where
\be\label{gauss}
 g(q,\Delta) &=& \frac{e^{-q^2/2\Delta^2}}{\sqrt{\Delta\sqrt{\pi}}},
\ee
and $\Delta$ is the width of the Gaussian and a measure of the degree of 
squeezing. $\Delta =1$ corresponds to the oscillator ground state, and
$\Delta<1$ indicates a squeezed state.
Using an ancilla qubit, initially in the zero state, $\ket{0}$,
the approximate encoded one state, $\ket{\wt{1}_1}$ is then created by
applying the sequence of operators,
\be\label{psig}
 \hat{H}e^{-i\alpha p_e\sigma_z}\hat{H},
\ee
where $\sigma_z$ is the Pauli $z$ matrix,
\be
 \sigma_z &=& \left(\begin{array}{cc}1 & 0 \\ 0 & -1 \end{array}\right),
\ee
applied to the qubit, and $\hat{H}$ is the Hadamard gate,
\be
 \hat{H} &=& \frac{1}{\sqrt2}
   \left(\begin{array}{cc} 1 & 1 \\ 1 & -1 \end{array}\right),
\ee
applied to the qubit.
Measuring the qubit in the zero state, which will occur with
probability 1/2, results in the continuous variable being left in the state,
\be\label{tilde1}
 \braket{q}{\wt{1}_1} &=& \frac{N}{\sqrt{2}}
     \left(g(q-\alpha,\Delta) + g(q+\alpha,\Delta)\right), 
\ee
where $N$ is a normalization factor, which is approximately equal to one, if
$\Delta/\alpha$ is small compared to one.
If the qubit is measured in the one state, the encoded variable is discarded
and we try again.
To create improved approximate encoded states, we iterate the following
procedure: \\
Given $\ket{\wt{1}_{n-1}}$, and a qubit in the state $\ket{0}$.
\begin{itemize}
\item Apply the operators:
\be\label{iterate}
\hat{H}e^{-i2^{n-1}\alpha p_e\sigma_z}\hat{H}
\ee
\item Measure the qubit. 
\item If the qubit is found in the state $\ket{0}$, then we have created
 $\ket{\wt{1}_n}$.
\item Else discard and start again.
\end{itemize}
Thus, with probability $1/2^n$, we create the approximate encoded state
\be\label{tilden}
 \braket{q}{\wt{1}_n} &=& \frac{N}{\sqrt{2^n}}
       \sum_{s=1}^{2^n} g(q+\alpha(1\!+\!2^n\!-\!2s),\Delta).
\ee
In momentum space the approximate encoded state has wave function
\be\label{p1n}
 \braket{p}{\wt{1}_n} &=&
 \left(\frac{\Delta}{2^n\sqrt{\pi}}\right)^{1/2} Ne^{-(p\Delta)^2/2}
  \frac{\sin{\alpha2^np}}{\sin{\alpha p}}.
\ee
\fig{approx1} depicts the approximate encoded state $\ket{\widetilde{1}_3}$,
with $\Delta = 0.15$ and $\alpha = \sqrt{\pi/2}$. This state will be generated 
with probability 1/8.
The approximate encoded zero state $\ket{\wt{0}_n}$ is created by displacing
the state $\ket{\wt{1}_n}$ by an amount $\alpha$ in the position variable.
Thus
\be\label{tild0n}
 \braket{q}{\wt{0}_n} &=&  
  \sum_{s=1}^{2^n} g(q+\alpha(2^n\!-\!2s),\Delta),
\ee
and
\be\label{p0n}
  \braket{p}{\wt{0}_n} &=& e^{-i\alpha p} \braket{p}{\wt{1}_n}.
\ee  
Because of the the $2^n$ term in \eqn{iterate}, the average energy of the
approximate encoded states will increase exponentially with $n$, however, as
we see in the following section, the probability of error decreases
exponentially with $n$.

It is perhaps also worth noting that alternative approximate encoded states, 
where the sign changes occur in position space rather than momentum space can 
be created by discarding the states when a $\ket{0}$ is measured instead of 
a $\ket{1}$.

\subsection{Fidelity of approximate encoded states}

As in \cite{Gottesman01}, the approximate encoded states $\ket{\wt{0}}$ and
$\ket{\wt{1}}$ will have negligible overlap if $\Delta$ is small compared to
$\alpha$.
In position space, the probability of mistaking an approximate encoded zero,
$\ket{\wt{0}}$, for an approximate encoded one, $\ket{\wt{1}}$ is simply the
probability of measuring the zero state nearer to an odd multiple of $\alpha$
than an even multiple.
The error probability will be bounded by the sum of each of the Gaussians'
tails,
\be
 \mbox{Error Prob} &<& 2^n \ 2 \int_{\alpha/2}^\infty dq 
 \ \left|\frac{g(q,\Delta)}{\sqrt{2^n}}\right|^2.
\ee 
Thus the error probability is independent of $n$, and using the asymptotic 
expansion of the error function,
\be\label{errorfun}
 \int_x^\infty dte^{-t^2} &=& \left(\frac{1}{2x}\right) e^{-x^2}
   \left[1 - O(1/x^2)\right],
\ee
it is not hard to show that error probability will be bounded by
\be
 \mbox{Error Prob} &<& \frac{4\Delta}{\sqrt{\pi}\alpha} 
   e^{-\frac{1}{8}\left(\frac{\alpha}{\Delta}\right)^2}.
\ee   
Therefore the likelihood of error becomes exponentially small for small
$\Delta/\alpha$.

In momentum space, we wish to determine the probability of finding
$(\ket{\wt{0}} - \ket{\wt{1}})/2$ closer to an even multiple of $\pi/\alpha$
than an odd multiple. Assuming $N \approx 1$, using \eqns{p1n}{p0n}, we
calculate the area under periodic part of the probability function,
\be
 \frac{|\braket{p}{\wt{0}_n} - \braket{1}{\wt{1}_n}|^2}{2}
\ee
about each even multiple of $\pi/\alpha$, divide this by the width
$2\pi/\alpha$, and multiple by the area of the Gaussian envelope,
\be
  \intd{p} e^{-(p\Delta)^2}.
\ee  
This gives a bound on the error probability of
\be
 \mbox{Error Prob} &<& \frac{1}{\pi 2^{n+1}},
\ee
which becomes exponentially small with $n$.

\section{Deterministic error recovery}\label{errorstates}

For robust quantum computation, it is necessary that our encoded states are 
comb-like in both the position and momentum quadratures. However, this is not
necessary for the ancilla systems used in error recovery. To correct an error 
in position, it is only necessary that the ancilla system is comb-like in 
position, and to correct an error in momentum it is only necessary that the 
ancilla system is comb-like in momentum.

\begin{figure}[ht]
 \centering
 \scalebox{0.7}{\includegraphics{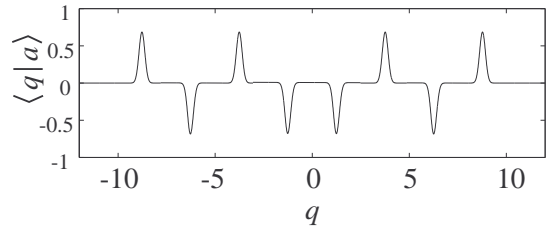}}
 \caption{Position wave function of an ancilla variable, $\ket{a}$, 
 which can be used in position quadrature error recovery.}
 \label{ancwave}
\end{figure}

This allows us to deterministically prepare ancilla systems for error recovery.
The ancilla system states can be prepared using the procedure described in 
\sect{approx}, except that we continue with the preparation procedure for
$n$ iterations, irrespective of whether the qubit is measured in the $\ket{0}$
or $\ket{1}$ state.
Thus, after three iterations, if the sequence of qubit measurements were say,
$\ket{1}, \ket{0}$ and $\ket{1}$, then we would be left with the state, 
$\ket{a}$, depicted in \fig{ancwave}. 
This state is no longer comb-like in momentum space, but it is still comb-like 
in position space. Thus, it could be used to perform position error recovery.
Likewise, ancilla variables appropriate for momentum quadrature error recovery
can be prepared by first squeezing the vacuum in the momentum quadrature, and 
replacing the qubit - continuous system interaction operator with
\be
  e^{-i2^{n-1}\alpha q_e\sigma_z}.
\ee

\section{Implementing in an ion trap}\label{iontrap}

There are several physical systems which enable a coupling between a continuous
quantum system and a discrete quantum system, such as a cavity QED system or an
ion trap. Here we discuss the possibility of creating approximate encoded 
states in an ion trap. 

Though scalable continuous variable quantum computation using ion traps seems
unlikely, the ion trap provides a good test bed for such first steps as creating
approximate encoded states, as the processes of decoherence within the ion trap
are well understood.

Consider a single $^9\mathrm{Be}^+$ ion, confined in a coaxial-resonator radio
frequency (RF)-ion trap, as described in \cite{Monroe96}, and references 
therein. The continuous quantum system is the vibrational mode of the ion, and
the two-level discrete system is the ground and first excited electronic
levels of the ion.

First it would be necessary to laser-cool the ion to the motional and electronic
ground state, as described in \cite{Monroe95}. Ideally, we would then need to 
squeeze the vibrational mode of the ion. This could prove a difficult task.
However, it is possible to create the sequence of operations described in 
\eqn{psig}.
The Hadamard operation is accomplished by a $\pi/2$-pulse, creating an equal
superposition of the ground and excited electronic states. A displacement beam
is then applied which excites the motion correlated to the excited state.
A $\pi$-pulse is then applied to exchange the internal states, and the 
displacement beam is applied again. Finally another $\pi/2$-pulse is applied,
executing the second Hadamard gate. 
The electronic level of the ion is then measured using another laser pulse, 
tuned to a transition between the first excited level and a higher level.
If fluorescence is observed, the ion has been measured in the $\ket{1}$ state.
The absence of fluorescence indicates that the ion is in the ground state.
In addition to the operations which we wish to implement, the ion trap system 
will undergo free evolution, so it will be necessary to couple the qubit, and
measure only once every period of oscillation.
In order to verify that the desired approximate encoded state had been created
it would then be necessary to carry out state tomography on the system.

\section{Conclusions}

For a quantum computer to become a reality, the daunting task of providing
adequate error correction needs to be fulfilled. At this point in time, it is
unclear which, if any, implementation scheme for quantum computation will
become viable. As the quantum mechanical oscillator is so prevalent in the
study of quantum mechanics, it appears to be a natural test bed for quantum
computation. Here we have shown how a continuous quantum system can be coupled
to a discrete two level quantum system in a manner which allows the continuous
quantum system to encode qubit. The ion trap provides a convenient
setting for this encoding scheme as it contains the required discrete and
continuous quantum variables.

\acknowledgments

BCT would like to thank R. Polkinghorne, G. Kociuba and P. Cochrane for helpful 
discussions.





\end{document}